\magnification =\magstep 1
\hsize = 14.5 truecm
\vsize = 23 truecm
\overfullrule 0pt
\hoffset 1truecm
\baselineskip=24pt
 
\newcount\refno
\refno=1
\def\y{\the\refno}
\def\myfoot#1{\footnote{$^{(\y)}$}{#1}
                 \advance\refno by 1}
\centerline{\bf Some Physical Consequences of Abrupt 
Changes in the}
\centerline{\bf Multipole Moments of a Gravitating 
Body}
\vskip 4truepc
\centerline{C. Barrab\`es\footnote*{E-mail: \tt barrabes@celfi.phys.univ-tours.fr} and G. F. Bressange\footnote{$\dagger$}{E-mail: \tt bressang@celfi.phys.univ-tours.fr}}
\centerline{Physics Department, UFR Sciences,}
\centerline{Universit\'e de Tours, 37200 France;}
\centerline{D\'epartment d'Astrophysique Relativiste et 
Cosmologie,}
\centerline{Observatoire de Paris,}
\centerline{92190 Meudon, France}
\vskip 1truepc
\centerline{and}
\vskip 1truepc
\centerline{P. A. Hogan\footnote{$\ddagger$}{E-mail: \tt phogan@ollamh.ucd.ie}}
\centerline{Mathematical Physics Department,}
\centerline{University College Dublin,}
\centerline{Belfield, Dublin 4, Ireland}
\vskip 3truepc
\noindent
PACS numbers: 04.30.+x, 04.20Jb
\vskip 1truepc
\noindent
The Barrab\`es--Israel theory of light-like shells 
in General Relativity is used to show explicitly 
that in general a light--like shell is accompanied 
by an impulsive gravitational wave. The gravitational 
wave is identified by its Petrov Type N contribution 
to a Dirac delta--function term in the Weyl conformal 
curvature tensor (with the delta--function singular 
on the null hypersurface history of the wave and 
shell). An example is described in which an asymptotically 
flat static vacuum Weyl space-time experiences a 
sudden change across a null hypersurface in the 
multipole moments of its isolated axially symmetric 
source. A light--like shell and an impulsive gravitational 
wave are identified, both having the null hypersurface 
as history. The stress--energy in the shell is 
dominated (at large distance from the source) by the jump 
in the monopole moment (the mass) of the source with 
the jump in the quadrupole moment mainly responsible for the 
stress being anisotropic. The gravitational wave owes 
its existence principally to the jump in the 
quadrupole moment of the source confirming what would 
be expected.

\vfill\eject
\noindent
{\bf 1. Introduction}
\vskip 1truepc
Very few exact solutions of Einstein's vacuum field 
equations exist describing gravitational waves from 
an isolated source having wave fronts homeomorphic to 
a 2--sphere (we will refer to such waves loosely as 
`spherical waves'). The principal example is the 
Robinson--Trautman [1] family of solutions. These are 
very special however because if the wave fronts are 
sufficiently smooth (free of conical singularities) and 
the field (Riemann tensor) contains no `wire' or 
`directional' singularities then the solutions approach 
a Schwarzschild limit exponentially in time [2, 3]. 
A limiting case is Penrose's [4] spherical impulsive 
gravitational wave propagating through flat space--time 
and similar solutions [5]. These all have the property 
that the curvature tensor of the space-time containing 
the history of the wave involves a Dirac delta function 
which is singular on the null hypersurface history 
of the wave and the coefficient of the delta function 
is singular along a generator of this null hypersurface 
(a wire singularity). The object of the present paper 
is to present an example of a wire singularity--free 
spherical impulsive gravitational wave propagating 
through a vacuum. To construct this we will use the 
Barrab\`es--Israel (BI)[6] theory of light--like 
shells in General Relativity. This is an extension 
to the null case of the usual extrinsic curvature 
technique [7] for studying non--null shells of matter. 
The Ricci tensor of the space--time in general exhibits 
a delta function behavior singular on the null hypersurface 
history of the shell (the coefficient of the delta 
function is constructed from the surface stress--energy 
tensor of the shell and this tensor is calculated 
using the BI approach).  
\vskip 1truepc 
We present in this paper two principal results: (1) If 
in general the Weyl tensor of the space--time containing 
the history of the light--like shell has a delta function 
singular on the history of the shell then we demonstrate 
explicitly, using the BI general theory, that the coefficient 
of this delta function splits into a matter part (due 
to anisotropic stresses in the shell) and a wave part 
describing an impulsive gravitational wave propagating 
in general independently of the shell. (2) Asymptotically 
flat solutions of the Weyl class of static axially 
symmetric vacuum gravitational fields are described 
by metric tensor components expressed as infinite series 
having coefficients involving the multipole moments 
of the isolated source [8]. To illustrate (1) we take 
a future directed null hypersurface which is 
asymptotically a future null--cone in the asymptotically 
flat Weyl space--time and assume a finite abrupt 
change takes place in these multipole moments across 
the null hypersurface. The surface stress--energy 
tensor concentrated on the null hypersurface 
is calculated at large distance along the null hypersurface 
from the history of the source and it is shown that the 
energy density of the shell, measured by a radially 
moving observer, is dominated by the jump in the 
monopole moment (the mass) of the source followed by 
terms proportional to the jumps in the dipole and 
quadrupole moments of the source. Jumps in higher 
multipole moments will contribute even less significantly 
and are not calculated here. There is an anisotropic 
surface stress dominated by the jump in the 
quadrupole moment. The delta function in the Weyl 
tensor has a coefficient with matter part and wave 
part, mentioned in (1) above, both of which are 
dominated by the jump in the quadrupole moment. In the 
case of the matter part this is due to the 
anisotropic stress in the shell while for the wave part it is a 
manifestation of the well--known property of gravitational 
waves from isolated sources that the lowest radiated 
multipole is the quadrupole. The Newman--Penrose 
components of the matter and wave parts of the 
coefficient of the delta function in the Weyl tensor are 
calculated on a null tetrad asymptotically parallel 
transported along the future--directed null geodesic generators 
of the null hypersurface. These components are non--
singular on any such generators 
and so {\it the Weyl tensor is free of wire 
singularities}. Since the null hypersurface chosen 
is asymptotically a future null--cone we can say 
that the impulsive gravitational wave accompanying 
the light--like shell is asymptotically spherical. 
\vskip 1truepc
The paper is organised as follows: In section 2 the 
BI technique is described as it applies to light--
like shells (which includes impulsive waves). The 
method given in BI is a unified treatment applicable 
to hypersurfaces which are non--null aswell as null 
and to non--null hypersurfaces which may become 
null in a limiting case. The identification of the 
matter and wave part of the coefficient of the 
delta function in the Weyl tensor is then presented. 
In section 3 the application to the asymptotically 
flat Weyl solutions is initiated by introducing these 
solutions, describing the transformation of the 
line--element to a form based on a family of 
null hypersurfaces $u={\rm constant}$ (say) (this is 
just the Bondi [9] form of the Weyl solutions; for 
completeness some of the original calculations by 
Bondi et al. [9] are summarised in the Appendix with 
comments relevant to the present application) and 
then describing how two such solutions with 
different multipole moments can be matched across 
one of the null hypersurfaces, $u=0$ (say). The 
physical properties of the boundary $u=0$ are worked 
out in section 4. Its interpretation as the history 
of both a light--like shell and an impulsive 
gravitational wave is verified asymptotically using 
the BI technique and the results derived in section 2. 
The paper ends with a discussion in section 5 commenting 
in particular on the absence of conical singularities 
on the wave front/shell and on the asymptotic behavior 
(`peeling' behavior) of the amplitude of the delta 
function in both the matter and wave parts of the 
Weyl tensor.
\vskip 4truepc
\noindent
{\bf 2. Barrab\`es--Israel Technique: The Null Case}
\vskip 1truepc
Consider space--time $M$ to be subdivided into two 
halves $M^+$ and $M^-$ each with boundary a null 
hypersurface $\Sigma$. Let $\{x^{\mu}_+\}$, with Greek 
indices taking values 1, 2, 3, 4,  be a 
local coordinate system in $M^+$ in terms of which 
the metric tensor components are $g^+_{\alpha\beta}$ 
and let $\{x^{\mu}_-\}$ be a local coordinate system 
in $M^-$ in terms of which the metric tensor components 
are $g^-_{\alpha\beta}$. Let $\{\xi ^a\}$, with Latin 
indices taking values 1, 2, 3, be local intrinsic 
coordinates on $\Sigma$ and the parametric equations 
of $\Sigma$ have the form $x^{\mu}_{\pm}=f^{\mu}_{\pm}
(\xi ^a)$ say. We thus have a basis of tangent vectors 
$e_a=\partial /\partial\xi ^a$ to $\Sigma$ and we assume 
that $M^+$ and $M^-$ are re--attached on $\Sigma$ in 
such a way that the induced metrics on $\Sigma$ from 
$M^+$ and $M^-$ match:
$$g_{ab}:=g_{\alpha\beta}\,e^\alpha _a\,e^\beta _b
\big |_{+}=g_{\alpha\beta}\,e^\alpha _a\,e^\beta _b
\big |_{-}\ ,\eqno(2.1)$$
where the colon followed by an equality sign denotes 
a definition, $e^{\alpha}_a\big |_{\pm}=\partial x^{\alpha}
_{\pm}/\partial\xi ^a$ are the components of the tangent 
vectors $e_a$ to $\Sigma$ evaluated on the $M^+$ side 
or $M^-$ side respectively. The symbol $\big |_{\pm}$ 
shall mean ``evaluated on the plus or minus side of 
$\Sigma$''. The manifold resulting from this re--
attachment of $M^+$ and $M^-$ on $\Sigma$ will be 
denoted by $M^+\cup M^-$. Let $n$ be normal to $\Sigma$ 
with components $n^{\mu}_{\pm}$ viewed on the plus or 
minus sides. Thus 
$$n^{\mu}\,n_{\mu}\big |_{\pm}=0\ ,\qquad {\rm and}\qquad 
n_{\mu}\,e^{\mu}_a\big |_{\pm}=0\ .\eqno(2.2)$$
Next choose a `transversal' $N$ on $\Sigma$ and require 
that its projection in the plus or minus sides of $\Sigma$ 
be the same (continuous) so that
$$N_{\mu}\,e^{\mu}_a\big |_{+}=N_{\mu}\,e^{\mu}_a\big |_{-}\ ,\eqno(2.3)$$
and, in addition, require
$$N_{\mu}\,N^{\mu}\big |_+=N_{\mu}\,N^{\mu}\big |_-\ .\eqno(2.4)$$
Finally define
$$\eta ^{-1}:=N_{\mu}\,n^{\mu}\ .\eqno(2.5)$$
Now the `transverse extrinsic curvature' of $\Sigma$ (a 
generalisation of the usual extrinsic curvature [7] to 
the null case) is defined by
$${\cal K}^{\pm}_{ab}:=-N_{\mu}\,e^{\mu}_{a;\nu}\,e^{\nu}
_b\big |_{\pm}={\cal K}^{\pm}_{ba}\ ,\eqno(2.6)$$
with the semi--colon denoting covariant differentiation. 
The jump in these components across $\Sigma$ is denoted 
by
$${1\over 2}\gamma _{ab}:=\left [{\cal K}_{ab}\right ]
:={\cal K}^+_{ab}-{\cal K}^-_{ab}\ ,\eqno(2.7)$$
and these quantities are independent of the choice 
of transversal (see [6], section II). We now extend 
$\gamma _{ab}$ to a four--tensor by padding--out with 
zeros (the only condition required on the extension 
$\gamma _{\mu\nu}$ being $\gamma _{\mu\nu}e^{\mu}_a\,
e^{\nu}_b=\gamma _{ab}$).
\vskip 1truepc
A calculation now of the Einstein tensor of the space--time 
$M^+\cup M^-$ results in general in an energy--momentum--
stress tensor which, in addition to parts ${}^+T^{\mu\nu}$ 
and ${}^-T^{\mu\nu}$ if $M^+$ and $M^-$ are non--vacuum, 
has a part concentrated on $\Sigma$ of the form [6]
$$T^{\mu\nu}=S^{\mu\nu}\,\delta (u)\ ,\eqno(2.8)$$
where the equation of $\Sigma$ is taken to be $u(x^{\mu})=0$, 
the normal to $\Sigma$ has the form $n_{\mu}=u_{,\mu}$, with 
the comma denoting partial differentiation, $\delta (u)$ is 
the Dirac delta function and $S^{\mu\nu}$ is given by [6]
$$16\pi\,\eta ^{-1}\,S^{\mu\nu}=2\gamma ^{(\mu}\,n^{\nu )}
-\gamma\,n^{\mu}\,n^{\nu}-\gamma ^{\dagger}\,g^{\mu\nu}
\ .\eqno(2.9)$$
Here round brackets denote symmetrisation and 
$$\gamma ^{\mu}:=\gamma ^{\mu\nu}\ ,\qquad \gamma ^{\dagger}:=
\gamma ^{\mu}\,n_{\mu}\ ,\qquad \gamma :=g^{\mu\nu}\,\gamma _{\mu\nu}
\ ,\eqno(2.10)$$
with the calculation of these quantities carried out on either side 
of $\Sigma$. We drop the plus or minus designation in such instances. 
The surface stress--energy tensor $S^{\mu\nu}$ in (2.9) has the 
property that
$$S^{\mu\nu}\,n_{\nu}=0\ ,\eqno(2.11)$$
and thus it can be expressed on the tangent basis $e_a$ as
$$S^{\mu\nu}=S^{ab}\,e^\mu _a\,e^\nu _b\ ,\eqno(2.12)$$
with (see [6])
$$16\pi\,\eta ^{-1}S^{ab}=2g_*^{c(a}\,l^{b)}\,\left (\gamma _{cd}\,
l^d\right )-g_*^{ab}\left (\gamma _{cd}\,l^c\,l^d\right )-
l^a\,l^b\,\left (g_*^{cd}\,\gamma _{cd}\right )\ .\eqno(2.13)$$
The three--vector $l^a$ is defined via the expansion 
$$n^{\mu}=l^a\,e^{\mu}_a\ ,\eqno(2.14)$$
remembering that since $\Sigma$ is null the normal to $\Sigma$ 
is also tangent to $\Sigma$. From (2.14) and the orthogonality 
of $n^{\mu}\ , e^\mu _a$ it follows that $g_{ab}\,l^b=0$. We 
note that since $\Sigma$ is null the induced metric with 
components $g_{ab}$ is degenerate or singular. In (2.13) $g_*^{ab}$ 
is a type of generalised inverse of $g_{ab}$ defined by [6]
$$g_*^{ac}\,g_{bc}=\delta ^a_b-\eta\,l^a\,N_b\ ,\eqno(2.15)$$
with $N_b:=N_{\mu}\,e^{\mu}_b$ which is continuous across $\Sigma$ 
by (2.3). The expression (2.15) determines $g_*^{ac}$ uniquely 
up to a multiple of $l^a\,l^c$. It is clear from (2.13) that if 
$\gamma _{ab}\,l^b=0$ then the surface stress--energy tensor is 
`isotropic' with 
$$16\pi\,\eta ^{-1}S^{ab}=-\left (g_*^{cd}\,\gamma _{cd}\right )\,
l^a\,l^b\ .\eqno(2.16)$$
This is equivalent, by (2.12) and (2.14), to $S^{\mu\nu}$ being 
proportional to $n^{\mu}\,n^{\nu}$.
\vskip 1truepc
The Weyl conformal curvature tensor of the space--time 
$M^+\cup M^-$ has in general a part concentrated on $\Sigma$ 
and given by [6]
$$C^{\kappa\lambda}{}_{\mu\nu}=\left\{2\eta\,n^{[\kappa}\,
\gamma ^{\lambda ]}{}_{[\mu}\,n_{\nu ]}-16\pi\,\delta ^{[\kappa}_{[
\mu}\,S^{\lambda ]}_{\nu ]}+{8\pi \over 3}S^{\alpha}_{\alpha}\,\delta 
^{\kappa\lambda}_{\mu\nu}\right\}\,\delta (u)\ .\eqno(2.17)$$
Here the square brackets denote skew--symmetrisation and 
$\delta ^{\kappa\lambda}_{\mu\nu}$ is the usual determinant of 
Kronecker deltas. It has been pointed out in [6] that there is 
a part of the first term in the coefficient of the delta function 
in (2.17) that is constructed from a part of $\gamma _{\mu\nu}$ 
which does not contribute to the surface stress--energy tensor. 
This part of (2.17) therefore is decoupled from the matter part. 
It describes an impulsive gravitational wave propagating with 
the shell and having $u=0$ as the history of its wave--front. 
To display this decomposition of (2.17) it is perhaps simplest 
to look first at the intrinsic form (2.13) for the stress--energy 
tensor. It is clear from (2.13) that a part of $\gamma _{ab}$, 
which we will denote by $\hat\gamma _{ab}$, does not contribute 
to the stress--energy tensor. This $\hat\gamma _{ab}$ satisfies 
$$\hat\gamma _{ab}\,l^b=0\ ,\qquad g^{ab}_*\,\hat\gamma _{ab}=0\ .
\eqno(2.18)$$
This means that $\hat\gamma _{ab}$ has two independent components 
(a fact which is related to the two degrees of freedom of 
polarisation in general present in the impulsive gravitational 
wave determined below by $\hat\gamma _{ab}$). The decomposition 
of (2.17) into wave and matter parts is best described by giving 
the components of these parts on the oblique basis $\left \{e^{\mu}_a\ , 
N^{\mu}\right\}$. Using (2.17) and hiving off the part of $\gamma _{ab}$ 
described above we find we can write
$$C_{\kappa\lambda\mu\nu}=\left (W_{\kappa\lambda\mu\nu}+
M_{\kappa\lambda\mu\nu}\right )\,\delta (u)\ .\eqno(2.19)$$
The components $W_{\kappa\lambda\mu\nu}\,e^\kappa _a\,e^\lambda _b
\,e^\mu _c\,e^\nu _d$ and $W_{\kappa\lambda\mu\nu}\,e^\kappa _a\,e^\lambda _b
\,e^\mu _c\,N^\nu $ of $W_{\kappa\lambda\mu\nu}$ vanish identically 
while
$$W_{\kappa\lambda\mu\nu}\,e^\kappa _a\,N^\lambda 
\,e^\mu _b\,N^\nu =-{1\over 2}\eta ^{-1}\hat\gamma _{ab}\ ,\eqno(2.20)$$
with
$$\hat\gamma _{ab}=\gamma _{ab}-{1\over 2}g^{cd}_*\,\gamma _{cd}\,
g_{ab}-2\eta\,l^d\,\gamma _{d(a}\,N_{b)}+\eta ^2\gamma _{cd}\,
l^c\,l^d\,N_a\,N_b\ .\eqno(2.21)$$
One easily checks that this $\hat\gamma _{ab}$ satisfies (2.18). 
Multiplying the components of $W_{\kappa\lambda\mu\nu}$ on the 
oblique basis by $l^a$ and using (2.14) and the first of (2.18) 
gives
$$W_{\kappa\lambda\mu\nu}\,n^\kappa =0\ ,\eqno(2.22)$$
so that this part of (2.19) is Type N in the Petrov classification 
with $n^\mu$ as four--fold degenerate principal null direction. 
Thus this part of the Weyl tensor in (2.19) describes an impulsive 
gravitational wave with propagation direction $n^\mu$ in space--time 
and with $u=0$ as the history of its wave--front. Now the 
components of the matter part of (2.19) on the oblique basis 
are
$$\eqalignno{M_{\kappa\lambda\mu\nu}\,e^\kappa _a\,e^\lambda _b\,
e^\mu _c\,e^\nu _d&=8\pi (g_{a[d}\,S_{c]b}-g_{b[d}\,S_{c]a})+
{16\pi\over 3}S^\alpha _\alpha \,g_{a[c}\,g_{d]b}\ ,&(2.23a)\cr
M_{\kappa\lambda\mu\nu}\,e^\kappa _a\,e^\lambda _b\,
e^\mu _c\,N^\nu &=-8\pi g_{c[a}\,e^\alpha _{b]}\,S_{\alpha\beta}\,
N^\beta +{16\pi\over 3}S^\alpha _\alpha\,g_{c[a}\,N_{b]}\ ,&(2.23b)\cr
M_{\kappa\lambda\mu\nu}\,e^\kappa _a\,N^\lambda \,
e^\mu _c\,N^\nu &=-8\pi S_{\alpha\beta}\,e^\alpha _{(a}\,N_{c)}\,
N^\beta +{4\pi\over 3}S^\alpha _\alpha\,N_a\,N_c\ .&(2.23c)\cr}$$
Here $S_{ab}:=S_{\mu\nu}\,e^\mu _a\,e^\nu _b$ which, of course, is not 
simply related to $S^{ab}$ in (2.13) by lowering indices 
because $g_{ab}$ is degenerate. We note that if $S^{\mu\nu}$ 
is isotropic in the sense described in (2.16), or in the sentence 
following (2.16), then $S_{ab}=0$ and in addition we see from 
(2.23) that $M_{\kappa\lambda\mu\nu}=0$ in this case. Finally 
we note that from the results of section III of [6] one can deduce 
that the matter part $M_{\kappa\lambda\mu\nu}$ is in general 
Petrov Type II and may specialise to Type III.
\vfill\eject
\noindent
{\bf 3. Asymptotically Flat Weyl Space--Times}
\vskip 1truepc
To illustrate the theory outlined in section 2 we 
consider the asymptotically flat Weyl static 
axially symmetric solutions of Einstein's vacuum 
field equations [8]. The line--element of these 
space--times has the form
$$ds^2=-R^2{\rm e}^{-2U}\,\left ({\rm e}^{2k}\,d\Theta ^2
+\sin ^2\Theta\,d\phi ^2\right )-{\rm e}^{2k-2U}\,
dR^2+{\rm e}^{2U}\,dt^2\ ,\eqno(3.1)$$
where $U, k$ are functions of the coordinates $\Theta , 
R$ given by the infinite series
$$\eqalignno{U&=\sum_{n=0}^{\infty}\,{a_n\over R^{n+1}}
\,P_n\ ,&(3.2a)\cr
k&=-\sum_{l, m=0}^{\infty}\,{a_l\,a_m\,(l+1)(m+1)\over 
l+m+2}\,{\left (P_l\,P_m-P_{l+1}\,P_{m+1}\over 
R^{l+m+2}\right )}\ ,&(3.2b)\cr}$$
where $a_n\ (n=0, 1, 2,\dots)$ are constants and 
$P_n=P_n\left (\cos\Theta\right )$ is the Legendre 
polynomial of degree $n$ in the variable $\cos\Theta$. 
The first few terms in the series (3.2a) may be written 
$$U=-{m\over R}-{D\,\cos\Theta\over R^2}-\left (
Q+{1\over 3}m^3\right )\,{P_2\left (\cos\Theta\right )
\over R^3}+\dots\ ,\eqno(3.3)$$
where, following Bondi et al. [9], we have written the 
constants $a_0, a_1, a_2$ in a form so that we can 
identify $m$ as the mass of the isolated source, $D$ 
as the dipole moment of the source and $Q$ as its 
quadrupole moment. The moments $D$ and $Q$ appear 
in (3.3) in such a way that if $D=Q=0$ then (3.3) 
represents the leading terms in the $1/R$--expansion 
of the Schwarzschild expression [8] for $U$. Corresponding 
to (3.3) we have for (3.2b)
$$k=-{m^2\over 2R^2}\,\sin ^2\Theta -{2m\,D\over R^3}\,\cos\Theta\,
\sin ^2\Theta +\dots\ .\eqno(3.4)$$
The line--element (3.1) with $U$ and $k$ in the forms 
(3.3) and (3.4) can be transformed to the Bondi form 
(the procedure for doing this, due to Bondi et al. [9], 
is outlined in the Appendix)
$$ds^2=-r^2\left\{f^{-1}d\theta ^2+f\,\sin ^2\theta\,d\phi ^2
\right\}+2g\,du\,dr+2h\,du\,d\theta +c\,du^2\ ,\eqno(3.5)$$
with
$$\eqalignno{f&=1-{Q\over r^3}\,\sin ^2\theta +
O\left (r^{-4}\right )\ ,&(3.6a)\cr
g&=1+O\left (r^{-4}\right )\ ,&(3.6b)\cr
h&={2D\over r}\,\sin\theta +{3Q\over r^2}\,\sin\theta
\,\cos\theta +O\left (r^{-3}\right )\ ,&(3.6c)\cr
c&=1-{2m\over r}-{2D\over r^2}\,\cos\theta -{Q\over r^3}
\left (3\cos ^2\theta -1\right )+O\left (r^{-4}\right )\ .&(3.6d)\cr}$$
For our present purpose it is useful to have solutions 
(3.1) expressed in a coordinate system based on a 
family of null hypersurfaces. In the form (3.5) the 
hypersurfaces $u={\rm constant}$ are {\it exactly} 
null (i.e. for all $r$ and not just for large $r$; this 
is clearly pointed out in the Appendix). Neglecting 
$O\left (r^{-4}\right )$--terms $u={\rm constant}$ 
are generated by the geodesic integral curves 
of the future--pointing null vector field $\partial/\partial r$ 
and $r$ is an affine parameter along them. These curves 
have expansion $\rho$ and shear $\sigma$ given by 
$$\eqalignno{\rho &={1\over r}+O\left (r^{-5}\right )\ ,&(3.7a)\cr
\sigma &={3Q\over 2r^4}\,\sin ^2\theta +O\left (r^{-5}\right )\ ,
&(3.7b)\cr}$$
demonstrating that for large values of $r$ (specifically, 
neglecting $O\left (r^{-4}\right )$--terms) the null 
hypersurfaces $u={\rm constant}$ in the space--time 
with line--element (3.5) are future null--cones.
\vskip 1truepc
To illustrate the theory described in section 2 we subdivide the 
space--time $M$ with line--element (3.5) into two 
halves $M^-$ and $M^+$ having $u=0$ (say) as common 
boundary. To the past of $u=0$, corresponding to $u<0$, 
the space--time $M^-$ is given by (3.5) and (3.6) 
with parameters $\left\{m_-, D_-, Q_-,\dots\right\}$ and 
coordinates $x^\mu _-=(\theta _-, \phi _-, r_-, u)$, 
while to the future of $u=0$, corresponding to $u>0$, 
the space--time $M^+$ is given by (3.5) and (3.6) with 
parameters $\left\{m_+, D_+, Q_+,\dots\right\}$ and 
coordinates $x^\mu _+=(\theta _+, \phi_+, r_+, u)$. 
We have taken $u_+=u_-=u$ here for convenience. To 
save on subscripts we shall 
henceforth drop the minus subscript on $x^\mu _-$ 
and on the parameters $\left\{m_-, D_-, Q_-,\dots\right\}$ and also use $\theta , \phi , r$ as intrinsic 
coordinates on $u=0$ (i.e. in the notation of 
section 2 we are putting $\xi ^a=(\theta , \phi , r)$). 
The line--element on $u=0$ induced from $M^+$ is
$$dl_+^2=-r_+^2\left\{f_+^{-1}\,d\theta _+^2+f_+
\,\sin ^2\theta _+\,d\phi ^2_+\right\}\ ,\eqno(3.8)$$
with 
$$f_+=1-{Q_+\over r_+^3}\,\sin ^2\theta _++O\left (r_+^{-4}\right )\ ,
\eqno(3.9)$$
while the induced line--element on $u=0$ from $M^-$ is  
$$dl^2=-r^2\left\{f^{-1}\,d\theta ^2+f
\,\sin ^2\theta \,d\phi ^2\right\}\ ,\eqno(3.10)$$
with 
$$f=1-{Q\over r^3}\,\sin ^2\theta +O\left (r^{-4}\right )\ .
\eqno(3.11)$$
We now re--attach $M^-$ and $M^+$ on $u=0$ requiring (3.8) and (3.10) 
to be the same line--element (see (2.1)) and this requirement gives 
the `matching conditions'
$$\eqalignno{\theta _+&=\theta +{\left [Q\right ]\over r^3}\,\sin\theta\,\cos\theta +
O\left (r^{-4}\right )\ ,&(3.12a)\cr
\phi _+&=\phi \ ,&(3.12b)\cr
r_+&=r+{\left [Q\right ]\over 2r^2}\,\left (1-3\cos ^2\theta\right )+O\left (
r^{-3}\right )\ ,&(3.12c)\cr}$$
with $\left [Q\right ]:=Q_+-Q$. From the perspective of the space--time 
$M^-\cup M^+$ we have an asymptotically flat Weyl solution $M^-$ 
undergoing an abrupt finite jump in its multipole moments across 
a null hypersurface $u=0$ resulting in the Weyl solution $M^+$. We 
now apply the BI theory, with special reference to the results of section 
2, to study the physical properties of the null hypersurface $u=0$.
\vskip 4truepc
\noindent
{\bf 4. Light--Like Shell and Gravitational Wave}
\vskip 1truepc
To apply the BI theory as outlined in section 2 we 
first calculate the tangent basis vectors $e_a=
\partial /\partial\xi ^a$ on the plus and minus sides of 
$\Sigma :\,u=0$. With $x^\mu _+$ given in terms of 
$x^\mu _-=x^\mu$ by (3.12) and with $\xi ^a=(\theta , 
\phi , r)$ we find that
$$e^\mu _a\big |_-=\delta ^\mu _a\ ,\eqno(4.1)$$
while
$$\eqalignno{e^\mu _1\,\big |_+&=\left [1+{\left [Q\right ]\over 
r^3}\,\cos 2\theta +O\left (r^{-4}\right ), 0, 
{3\left [Q\right ]\over 2r^2}\,\sin 2\theta +O\left (
r^{-3}\right ), 0\right ]\ ,&(4.2a)\cr
e^\mu _2\,\big |_+&=(0, 1, 0, 0)\ ,&(4.2b)\cr
e^\mu _3\,\big |_+&=\left [{3\left [Q\right ]\over 2r^2}\,\sin 2\theta 
+O\left (r^{-3}\right ), 0, 1-{\left [Q\right ]\over r^3}\,
(1-3\cos ^2\theta )+O\left (r^{-4}\right ), 0\right ]\ .
&(4.2c)\cr}$$
The normal to $u=0$ is
$$n_\mu\,dx^\mu\,\big |_{\pm}=du\ .\eqno(4.3)$$    
As transversal on the minus side we can take
$${}^{-}N_{\mu}=\left [0, 0, 1, {1\over 2}-{m\over r}
-{D\over r^2}\,\cos\theta -{Q\over 2r^3}\,\left (3\cos ^2\theta 
-1\right )+O\left (r^{-4}\right )\,\right ]\ ,\eqno(4.4)$$     
and thus when viewed on the plus side we find, after 
insisting on (2.3) and (2.4) being satisfied, that the 
components of the transversal are ${}^+N_\mu$ with 
$$\eqalignno{{}^+N_1&={3\left [Q\right ]\over 2r^2}\,\sin 2\theta 
+O\left (r^{-3}\right )\ ,&(4.5a)\cr
{}^+N_2&=0\ ,&(4.5b)\cr
{}^+N_3&=1+{\left [Q\right ]\over r^3}\,\left (1-3\cos ^2\theta \right )
+O\left (r^{-4}\right )\ ,&(4.5c)\cr
{}^+N_4&={1\over 2}-{m_+\over r}-{D_+\over r^2}\,\cos\theta 
+{\left (2Q_+-Q\right )\over 2r^3}\,\left (1-3\cos ^2\theta \right )\cr
&+O\left (r^{-4}\right )\ .&(4.5d)\cr}$$
Now $\eta$ given by (2.5) has the form
$$\eta =1+O\left (r^{-3}\right )\ .\eqno(4.6)$$
The transverse extrinsic curvature on the plus and 
minus sides of $u=0$ is calculated now from (2.6). On 
the minus side we find
$$\eqalignno{{\cal K}^-_{11}&={r\over 2}-m-{3D\over r}\,
\cos\theta +{Q\over 4r^2}\,\left (13-29\,\cos ^2\theta\right )
+O\left (r^{-3}\right )\ ,&(4.7a)\cr
{\cal K}^-_{12}&=0\ ,&(4.7b)\cr
{\cal K}^-_{22}&={r\over 2}\sin ^2\theta -m\,\sin ^2\theta 
-{3D\over r}\,\cos\theta\,\sin ^2\theta \cr
&+{Q\over 4r^2}\,\left (3-19\,\cos ^2\theta\right )
\,\sin ^2\theta +O\left (r^{-3}\right )\ ,&(4.7c)\cr
{\cal K}^-_{13}&={3D\over r^2}\,\sin\theta +
{6Q\over r^3}\,\sin\theta\,\cos\theta +O\left (r^{-4}\right )\ ,
&(4.7d)\cr
{\cal K}^-_{23}&=0\ ,&(4.7e)\cr
{\cal K}^-_{33}&=O\left (r^{-5}\right )\ .&(4.7f)\cr}$$
On the plus side we have
$$\eqalignno{{\cal K}^+_{11}&={r\over 2}-m_+-{3D_+\over r}\,
\cos\theta +{Q_+\over 4r^2}\,\left (13-29\,\cos ^2\theta\right )\cr
&+{\left [Q\right ]\over 4r^2}\,\left (11-25\,\cos ^2\theta\right )
+O\left (r^{-3}\right )\ ,&(4.8a)\cr
{\cal K}^+_{12}&=0\ ,&(4.8b)\cr
{\cal K}^+_{22}&={r\over 2}\sin ^2\theta -m_+\,\sin ^2\theta 
-{3D_+\over r}\,\cos\theta\,\sin ^2\theta\cr 
&+{Q_+\over 4r^2}\,\left (3-19\,\cos ^2\theta\right )
\,\sin ^2\theta +{\left [Q\right ]\over 4r^2}\,\left (3+7\,\cos ^2
\theta\right )\sin ^2\theta
+O\left (r^{-3}\right )\ ,&(4.8c)\cr
{\cal K}^+_{13}&={3\left [Q\right ]\over 4r}\,\sin 2\theta +
O\left (r^{-2}\right )\ ,
&(4.8d)\cr
{\cal K}^+_{23}&=0\ ,&(4.8e)\cr
{\cal K}^+_{33}&=O\left (r^{-3}\right )\ .&(4.8f)\cr}$$
From (4.7) and (4.8) the jump $\gamma _{ab}$ in 
the transverse extrinsic curvature defined by (2.7) 
has the following components:
$$\eqalignno{\gamma _{11}&=-2[m]-{6\left [D\right ]\over 
r}\,\cos\theta +{\left [Q\right ]\over r^2}\,(12-27\,\cos ^2\theta 
)+O\left (r^{-3}\right )\ ,&(4.9a)\cr
\gamma _{22}&=-2[m]\,\sin ^2\theta -{6\left [D\right ]
\over r}\,\cos\theta\,\sin ^2\theta\cr
& +{3\left [Q\right ]
\over r^2}\,(1-2\,\cos ^2\theta )\,\sin ^2\theta +
O\left (r^{-3}\right )\ ,&(4.9b)\cr
\gamma _{12}&=0\ ,\qquad \gamma _{23}=0\ ,&(4.9c)\cr
\gamma _{13}&={3\left [Q\right ]\over r}\,\sin\theta\,
\cos\theta +O\left (r^{-2}\right )\ ,&(4.9d)\cr
\gamma _{33}&=\left (r^{-3}\right )\ ,&(4.9e)\cr}$$
where, inkeeping with earlier notation, we have put $[m]:=
m_+-m$ and $\left [D\right ]:=D_+-D$. Finally $\hat\gamma _{ab}$ 
in (2.21) turns out to have all but two components vanishing 
identically. These two components are
$$\eqalignno{\hat\gamma _{11}&={1\over 2}\left (\gamma _{11}-
\csc ^2\theta\,\gamma _{22}\right )+O\left (r^{-3}\right )\ ,&(4.10a)\cr
\hat\gamma _{22}&=-\hat\gamma _{11}\,\sin ^2\theta +O\left (r^{-3}\right )
\ .&(4.10b)\cr}$$
Now the leading terms in the surface stress--energy tensor $S^{\mu\nu}$ 
are calculated from (2.9) or (2.13). We find $S^{\mu 4}=S^{12}=
S^{23}=0$ while
$$\eqalignno{S^{11}&=O\left (r^{-5}\right )\ ,\qquad S^{22}=
O\left (r^{-5}\right )\ ,&(4.11a)\cr
16\pi\,S^{13}&=-{3\left [Q\right ]\over r^{3}}\,\sin\theta\,
\cos\theta +O\left (r^{-4}\right )\ ,&(4.11b)\cr
16\pi\,S^{33}&=-{4[m]\over r^2}-{12\left [D\right ]\over r^3}\,
\cos\theta +{3\left [Q\right ]\over r^4}\,(5-11\,\cos ^2\theta )
+O\left (r^{-5}\right ).&(4.11c)\cr}$$
Thus the stress in the light--like shell with history $u=0$ is 
anisotropic due primarily (for large $r$) to the jump in 
the quadrupole moment $\left [Q\right ]$ (on account of (4.11b)). 
The surface energy density of the shell measured by a radially 
moving observer (see [6]) is, by (4.11c), a positive multiple 
of 
$$\sigma :=-{1\over 4\pi r^2}\left\{[m]+{3\left [D\right ]\over r}
\,
\cos\theta -{3\left [Q\right ]\over 4r^2}\,(5-11\,\cos ^2\theta )
+O\left (r^{-3}\right )\right\}\ .\eqno(4.12)$$
This is dominated by the jump in the monopole moment (the mass 
of the source). It would be natural to assume that $[m]<0$ so 
that the source suffers a mass--loss.
\vskip 1truepc
It is convenient to define, in view of the line--element (3.5), 
a null tetrad
$$\eqalignno{M^\mu&=\left (-{1\over r\sqrt{2}}\,f^{1/2}, -{i\over r
\sqrt{2}\,\sin\theta}\,f^{-1/2}, 0, 0\right )\ ,&(4.13a)\cr
\bar M^\mu&=\left (-{1\over r\sqrt{2}}\,f^{1/2}, {i\over r
\sqrt{2}\,\sin\theta}\,f^{-1/2}, 0, 0\right )\ ,&(4.13b)\cr
n^\mu&=\left (0, 0, 1+O\left (r^{-4}\right ), 0\right )\ ,&(4.13c)\cr
N^\mu&=\left (O\left (r^{-3}\right ), 0, -{c\over 2}, 1\right )\ ,
&(4.13d)\cr}$$
with $f, c$ given by (3.6a, d) respectively. This tetrad is 
asymptotically parallel transported along the integral curves 
of $\partial/\partial r$. On this tetrad the Newman--Penrose 
components of the matter part of the delta function (2.19) in 
the Weyl tensor (which we denote by ${}^M\Psi _A$ with $A=0, 1, 2, 
3, 4$) are given, using (2.23), by
$$\eqalignno{{}^M\Psi _0&=0\ ,\qquad {}^M\Psi _1=0\ ,\qquad 
{}^M\Psi _2=O\left (r^{-3}\right )\ ,&(4.14a)\cr
{}^M\Psi _3&={9\left [Q\right ]\over 4\sqrt {2}\,r^2}\,\sin\theta\,
\cos\theta +O\left (r^{-3}\right )\ ,&(4.14b)\cr
{}^M\Psi _4&=O\left (r^{-5}\right )\ .&(4.14c)\cr}$$
This is thus predominantly Petrov Type III (with $n^\mu$ as 
generate principal null direction) due to the anisotropy in 
the stress (4.11b) (which in turn is due to $\left [Q\right ]\neq 0$). 
The leading term in (4.14b) has clearly no singularity for 
$0\leq\theta\leq\pi$ (and thus no wire singularity). All Newman--
Penrose components on the null tetrad (4.13) of the wave part of 
the delta function (2.19) in the Weyl tensor (which we denote 
by ${}^W\Psi _A$) vanish with the exception, calculated from (2.20) 
and (4.13), of
$${}^W\Psi _4=-{3\left [Q\right ]\over 4r^4}\,(3-7\,\cos ^2\theta )
+O\left (r^{-5}\right )\ .\eqno(4.15)$$
This impulsive gravitational wave clearly owes its existence 
primarily to the jump in the quadrupole moment of the source 
across $u=0$ and also is manifestly free of wire singularities.
\vfill\eject
\noindent
{\bf 5. Discussion}
\vskip 1truepc
The degenerate metric induced on $u=0$, the history of an outgoing 
light--like shell and of an impulsive gravitational wave (as has 
been verified asymptotically using the BI technique in section 4), 
is given asymptotically by (3.8) and (3.10). The line--element (3.10) 
can be written, putting $\cos\theta =x$, as
$$dl^2=-r^2\,\left\{G^{-1}dx^2+G\,d\phi ^2\right\}\ ,\eqno(5.1)$$
with
$$G=(1-x^2)\,\left\{1-{Q(1-x^2)\over r^3}+O\left (r^{-4}\right )\right \}\ 
,\eqno(5.2)$$
which, for each $r$, is a standard form for the line--element on a 
2--surface of revolution embedded in three dimensional Euclidean 
space (see, for example [10]) with $-1\leq x\leq +1$ and $0\leq\phi <2\pi$. 
The Gaussian curvature is $K/r^2$ with 
$$K=-{1\over 2}G''=1+{4Q\over r^3}\,P_2(x)+O\left (r^{-4}\right )\ .\eqno(5.3)$$
Here the prime denotes differentiation with respect to $x$. Neglecting 
$O\left (r^{-4}\right )$--terms, we see that $G'(+1)+G'(-1)=0$ 
and this together with $\phi$ ranging from $0$ to $2\pi$ means (see [10]) 
that there are no conical singularities at the north or south poles 
of the 2--surface. In fact, by the Gauss--Bonnet theorem it is clear 
from the form of the line--element (5.1) and $K$ in (5.3) that neglecting 
$O\left (r^{-4}\right )$--terms the 2--surface is topologically 
spherical. Hence the light--like shell and the impulsive gravitational 
wave can be considered asymptotically spherical in this sense.
\vskip 1truepc
Finally with ${}^M\Psi _3=O\left (r^{-2}\right )$ and ${}^W\Psi _4=
O\left (r^{-4}\right )$ we see an unfamiliar Peeling behavior. This 
is due to (a) the conventional Peeling behavior occuring asymptotically 
in the field of an isolated source {\it with history confined to a 
time--like world--tube} of compact cross--section whereas the source 
of ${}^M\Psi _A$ and ${}^W\Psi _A$ is a light--shell and a wave 
with the null hypersurface $u=0$ as history in space--time and 
(b) since in our case 
the radiation part of the field is in direct competition with the 
matter part it is no surprise that, in terms of $r^{-1}$, the 
amplitude of the matter part dominates that of the radiation part.
\vskip 4truepc
\noindent
{\bf References}
\vskip 1truepc
\noindent\item{[1]} I. Robinson and A. Trautman, {\it Proc. R. Soc. 
London} A{\bf 265}, 463 (1962).
\vskip 1truepc
\noindent\item{[2]} B. Lukacs, Z. Perj\`es, J. Porter and A. Sebestyen, 
{\it Gen. Rel. Grav.} {\bf 16}, 691 (1984).
\vskip 1truepc
\noindent\item{[3]} I. Robinson, {\it Class. Quantum Grav.} {\bf 6}, 
1863 (1989).
\vskip 1truepc
\noindent\item{[4]} R. Penrose, in {\it General Relativity} (papers 
in honour of J. L. Synge), edited by L. O' Raifeartaigh (Clarendon 
Press, Oxford, 1972), p.101.
\vskip 1truepc
\noindent\item{[5]} G. A. Alekseev and J. B. Griffiths, {\it Class. 
Quantum Grav.} {\bf 13}, L13 (1996).
\vskip 1truepc
\noindent\item{[6]} C. Barrab\`es and W. Israel, {\it Phys. Rev.} 
D{\bf 43}, 1129 (1991).
\vskip 1truepc
\noindent\item{[7]} W. Israel, {\it Nuovo Cimento} B{\bf 44}, 1 (1966).
\vskip 1truepc
\noindent\item{[8]} D. Kramer, H. Stephani, E. Herlt and M. A. H. MacCallum, 
{\it Exact Solutions of Einstein's Field Equations} (Cambridge University 
Press, Cambridge, 1980), p.201.
\vskip 1truepc
\noindent\item{[9]} H. Bondi, M. G. J. van der Burgh and A. W. K. 
Metzner, {\it Proc. R. Soc. London} A{\bf 269}, 21 (1962).
\vskip 1truepc
\noindent\item{[10]} I. Robinson and J. R. Robinson in {\it General Relativity} (papers 
in honour of J. L. Synge), edited by L. O' Raifeartaigh (Clarendon 
Press, Oxford, 1972), p.151.  
\vskip 4truepc
\centerline{\bf Appendix}
\vskip 1truepc
\centerline{Transformation of Weyl Solution to 
Bondi Form}
\vskip 2truepc\noindent
To make the present paper as self contained as 
possible we briefly outline and discuss here the 
transformation of the Weyl solution given by (3.1, 2) 
to the Bondi [9] form (3.5, 6). This transformation is 
given in Appendix 4 of [9]. We wish to emphasise 
aspects of the procedure which are particularly 
relevant to the topic under consideration in the 
present paper.
\vskip 1truepc
Starting with (3.1) the coordinate transformation 
$$\eqalignno{t&=u+F\left (R, \theta\right )\ ,&(1)\cr
\Theta &=\Theta\left (R, \theta\right )\ ,&(2)\cr}$$
is made with the functions  $F, \Theta$ chosen so that 
no $dr\,d\theta$ or $dr^2$ terms appear in the line--
element. This will be achieved provided $F, \Theta$ 
satisfy the partial differential equations
$$\eqalignno{{\rm e}^{2U}F_R\,F_{\theta}&=R^2\,
{\rm e}^{2k-2U}\Theta _R\,\Theta _{\theta}\ ,&(3)\cr
{\rm e}^{2U}F_R^2&={\rm e}^{2k-2U}\,\left (1+R^2\,
\Theta _R^2\right )\ ,&(4)\cr}$$
with the suBIcripts on $F, \Theta$ indicating partial 
derivatives with respect to $R, \theta$ as appropriate.
At this point the line--element reads
$$ds^2=\left (-R^2{\rm e}^{2k-2U}\Theta ^2_{\theta}
+{\rm e}^{2U}\,F^2_{\theta}\right )\,d\theta ^2-
R^2\,{\rm e}^{-2U}\sin ^2\Theta\,d\phi ^2$$
$$+2{\rm e}^{2U}F_R\,du\,dR+2{\rm e}^{2U}F_{\theta}\,
du\,d\theta +{\rm e}^{2U}\,du^2\ .\eqno(5)$$
We emphasise now that the hypersurfaces $u={\rm constant}$ 
are {\it null}. Using $F_{R\theta}=F_{\theta R}$ and 
$\Theta _{R\theta}=\Theta _{\theta R}$ in (3), (4) we 
can, following [9], eliminate $F$ from (3), (4) and 
arrive at  
$$\Theta _\theta\,\left [{R^4{\rm e}^{2k-4U}\Theta ^2_R\over 1+R^2\,\Theta 
^2_R}\right ]_R=R^2\left ({\rm e}^{2k-4U}\right )
_{\theta}\,\Theta _R\ .\eqno(6)$$
This equation is now solved approximately for large 
values of $R$ by (see [9])
$$\Theta =\theta +{p'\over 4R^2}+{1\over R^3}
\left ({1\over 12}q'-{1\over 2}mp'\right )+\dots\ ,\eqno(7)$$
with
$$\eqalignno{p&=4D\cos\theta +m^2(7+\cos ^2\theta )\ ,&(8)\cr
q&=2Q(3\cos ^2\theta -1)+4m\,D\,\cos\theta\,(3+\cos ^2\theta )
+6m^3\,(1+\cos ^2\theta )\ ,&(9)\cr}$$
and the primes in (7) on $p, q$ indicating derivatives 
with respect to $\theta$. Now $F$ is obtained from 
(3), (4) as
$$F=R+2m\,{\rm log}R+{1\over R}(2m^2-{1\over 2}p)
+{1\over R^2}(-2m^3+{1\over 2}m\,p-{1\over 4}q)
+\dots\ .\eqno(10)$$
Finally we make the transformation $R=R(r, \theta )$ 
given by
$$R^2{\rm e}^{-2U}\,\sin ^2\Theta\,\left (R^2{\rm e}
^{2k-2U}\Theta ^2_{\theta}-{\rm e}^{2U}\,F^2_{\theta}
\right )=r^4\,\sin ^2\theta\ .\eqno(11)$$
This leads, for large values of $r$, to 
$$R=r-m-{m^2\over 2r}\sin ^2\theta -{m\over 2r^2}
(2D\,\cos\theta +m^2)\,\sin ^2\theta +\dots\ .\eqno(12)$$
Putting this in $F$ given by (10) and $\Theta$ given by (7) 
we construct the transformation (1), (2) for large $r$ 
leading from (3.1,2) to (3.5,6). We emphasise that although 
the differential equations (3), (4) for $F, \Theta$ 
and the algebraic equation (11) for $R$ have been 
solved only for large values of $r$, the null 
hypersurfaces $u={\rm constant}$ are exactly null 
(for all values of $r$).

\bye